\documentclass[aip,jcp,reprint,noshowkeys,superscriptaddress]{revtex4-1}
\usepackage{graphicx,dcolumn,bm,xcolor,microtype,multirow,amscd,amsmath,amssymb,amsfonts,physics,wrapfig,bbold,siunitx,xspace,soul}
\usepackage[version=4]{mhchem}

\usepackage[utf8]{inputenc}
\usepackage[T1]{fontenc}

\usepackage{hyperref}
\hypersetup{
    colorlinks,
    linkcolor={red!50!black},
    citecolor={red!70!black},
    urlcolor={red!80!black}
}

\usepackage{listings}
\definecolor{codegreen}{rgb}{0.58,0.4,0.2}
\definecolor{codegray}{rgb}{0.5,0.5,0.5}
\definecolor{codepurple}{rgb}{0.25,0.35,0.55}
\definecolor{codeblue}{rgb}{0.30,0.60,0.8}
\definecolor{backcolour}{rgb}{0.98,0.98,0.98}
\definecolor{mygray}{rgb}{0.5,0.5,0.5}

\definecolor{sqred}{rgb}{0.85,0.1,0.1}
\definecolor{sqgreen}{rgb}{0.25,0.65,0.15}
\definecolor{sqorange}{rgb}{0.90,0.50,0.15}
\definecolor{sqblue}{rgb}{0.10,0.3,0.60}

\lstdefinestyle{mystyle}{
    backgroundcolor=\color{backcolour},
    commentstyle=\color{codegreen},
    keywordstyle=\color{codeblue},
    numberstyle=\tiny\color{codegray},
    stringstyle=\color{codepurple},
    basicstyle=\ttfamily\footnotesize,
    breakatwhitespace=false,
    breaklines=true,
    captionpos=b,
    keepspaces=true,
    numbers=left,
    numbersep=5pt,
    numberstyle=\ttfamily\tiny\color{mygray},
    showspaces=false,
    showstringspaces=false,
    showtabs=false,
    tabsize=2
  }

  \newcolumntype{d}{D{.}{.}{-1}}

\newcommand{\mc}{\multicolumn}

\newcommand{\ra}{\rightarrow}
\newcommand{\pis}{\pi^\star}
\newcommand{\npi}{n\ra\pis}
\newcommand{\ppi}{\pi\ra\pis}

\newcommand{\lmax}{\lambda_{\text{max}}}
\newcommand{\eoo}{E_{\text{0-0}}}

\newcommand{\Pop}{6-31+G(d)}
\newcommand{\AVDZ}{{aug}-cc-pVDZ}
\newcommand{\AVTZ}{{aug}-cc-pVTZ}

\newcommand{\LCPQ}{Laboratoire de Chimie et Physique Quantiques (UMR 5626), Universit\'e de Toulouse, CNRS, UPS, France}
\newcommand{\CEISAM}{Nantes Universit\'e, CNRS,  CEISAM UMR 6230, F-44000 Nantes, France}
\newcommand{\IUF}{Institut Universitaire de France (IUF), F-75005 Paris, France}

\begin{document}	

\title{A Mountaineering Strategy to Excited States: Accurate Vertical Transition Energies and Benchmarks for Substituted Benzenes}
\author{Pierre-Fran\c{c}ois \surname{Loos}}
	\affiliation{\LCPQ}
\author{Denis \surname{Jacquemin}}
	\email{Denis.Jacquemin@univ-nantes.fr}
	\affiliation{\CEISAM}
	\affiliation{\IUF}

\begin{abstract}
In an effort to expand the existing QUEST database of accurate vertical transition energies [\href{https://doi.org/10.1002/wcms.1517}{V\'eril et al.~\textit{WIREs Comput.~Mol.~Sci.} \textbf{2021}, \textit{11}, e1517}], 
we have modeled more than 100 electronic excited states of different natures (local, charge-transfer, Rydberg, singlet, and triplet) in a dozen of mono- and di-substituted benzenes, including aniline, benzonitrile, 
chlorobenzene, fluorobenzene, nitrobenzene, among others.  To establish theoretical best estimates for these vertical excitation energies, we have employed advanced coupled-cluster 
methods including iterative triples (CC3 and CCSDT) and, when technically possible, iterative quadruples (CC4). These high-level computational approaches provide a robust foundation for benchmarking a series of 
popular wave function methods. The evaluated methods all include contributions from double excitations (ADC(2), CC2, CCSD, CIS(D), EOM-MP2, STEOM-CCSD), along with schemes that also incorporate perturbative or iterative triples
(ADC(3), CCSDR(3), CCSD(T)(a)$^\star$, and CCSDT-3). This systematic exploration not only broadens the scope of the QUEST database but also facilitates a rigorous assessment of different theoretical approaches
in the framework of a homologous chemical series, offering valuable insights into the accuracy and reliability of these methods in such cases. We found that both ADC(2.5) and CCSDT-3 can provide very consistent
estimates, whereas among less expensive methods SCS-CC2 is likely the most effective approach. Importantly, we show that some lower order methods may offer reasonable trends in the homologous series while providing
quite large average errors, and \emph{vice versa}. Consequently, benchmarking the accuracy of a model based solely on absolute transition energies may not be meaningful for applications involving a series of similar compounds.
  
\bigskip
\begin{center}
	\boxed{\includegraphics[width=0.4\linewidth]{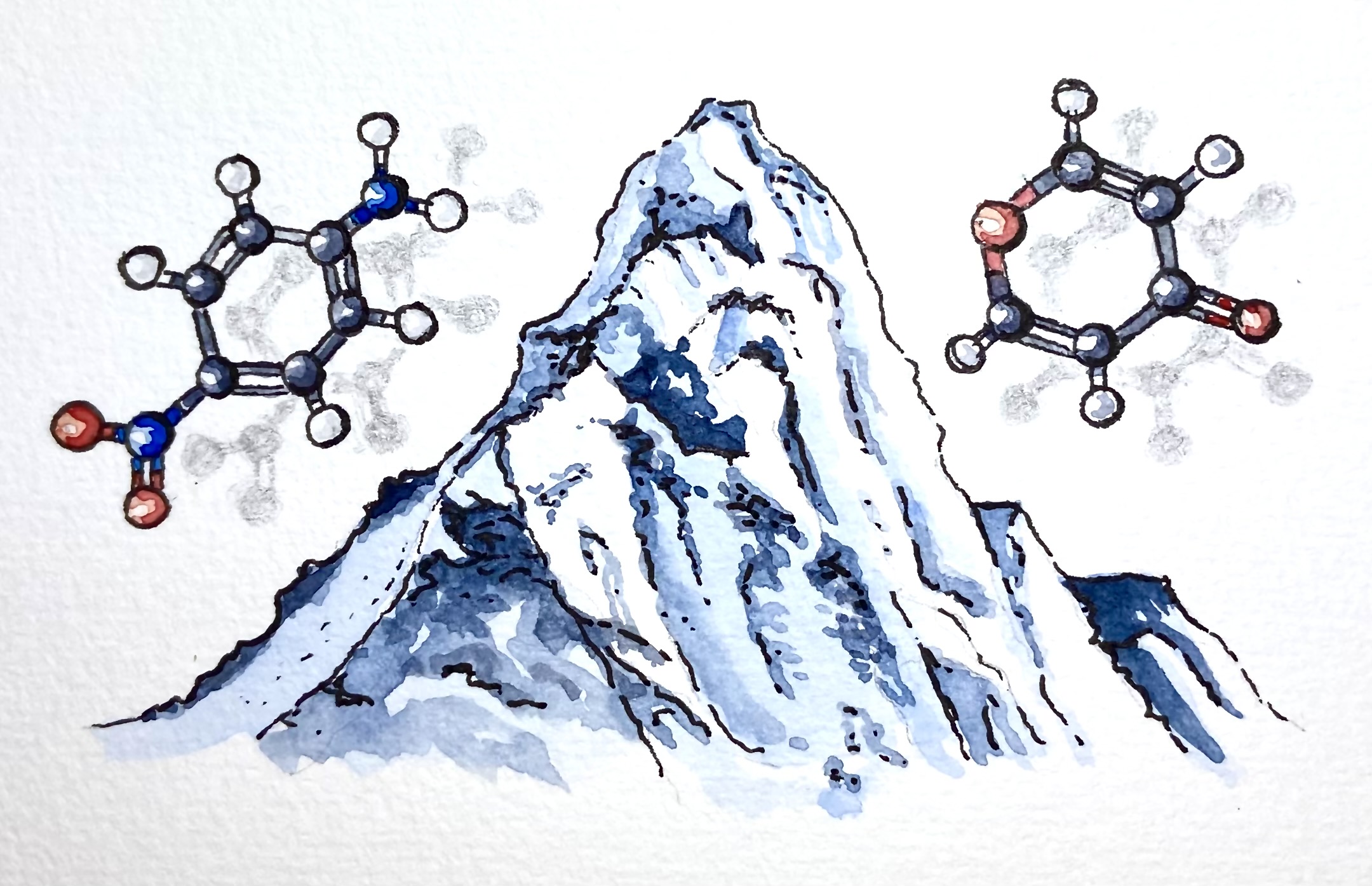}}
\end{center}
\bigskip
\end{abstract}

\maketitle

\section{Introduction}
\label{sec:Intro}

The efficient harnessing of photophysical and photochemical phenomena still presents a formidable challenge for chemists. These intricate processes unfold at an ultrafast timescale and often entail the  
interplay of multiple electronic states, introducing a level of complexity beyond the scope of traditional thermal chemistry. In response to this challenge, the experimental and theoretical communities have 
actively pursued the development of innovative methodologies, aiming to characterize electronic excited states (EES), which serve as central objects for comprehending and rationalizing interactions 
between light and matter.

On the theoretical side, efforts have been dedicated to developing efficient and computationally economical methodologies offering chemically relevant descriptions of EES. \cite{Loo20c} In particular, time-dependent density-functional 
theory (TD-DFT),  \cite{Run84,Cas95,Ull12b} a relatively recent theoretical approach, has emerged as a powerful tool for modeling optical spectra, probing the nature of the EES, and getting estimates of their properties.\cite{Ada13a,Lau14} However, 
like all theoretical models, TD-DFT comes with an intrinsic error, and the quantification of its magnitude is required, yet challenging. To this end, accurate reference values are essential. In the case of EES, relying on experimental data as
references is far from straightforward. Indeed, the measurements of rather simple EES properties, such as geometries and dipoles, exhibit significant uncertainties and are only possible for compact systems. Conversely, highly accurate 
measurements of transition energies are available, albeit primarily for 0-0 (non-vertical) transitions, thus making the direct comparison between the experimental and theoretical data more difficult. \cite{Loo19b} Consequently, the development 
of comprehensive databases of theoretical vertical transition energies (VTEs) remains necessary to assess the performances of more computational affordable methods dedicated to EES.

Inspired by the seminal works of Thiel's group, \cite{Sch08,Sau09,Sil10b,Sil10c} our joint efforts, started in 2018, have resulted in the establishment of an extensive database of reference VTEs, known as QUEST. \cite{Ver21} A summary 
of our contributions during the 2018--2021 period can be found in Ref.~\citenum{Ver21}, which gathers approximately 500 VTE entries, the majority of which are deemed chemically accurate, with deviations not larger than 0.05 eV 
from the exact  values obtained through full configuration interaction (FCI) calculations. To achieve this, we relied on two theoretical models: selected configuration interaction (SCI), an elegant and effective method to quickly approach the 
FCI limit, but limited in practice to quite  small molecules, \cite{Hur73,Gin13,Gin15,Sch16,Hol17,Gar18,Chi18,Eri21} and coupled-cluster (CC) theory, which offers a well-defined path for systematic improvement with the following sequence of 
methods: CC2,  \cite{Chr95,Hat00} CCSD,  \cite{Pur82,Scu87,Koc90b,Sta93,Sta93b} CC3, \cite{Chr95b,Koc95,Koc97}  CCSDT,  \cite{Nog87,Scu88,Kuc01,Kow01,Kow01b} CC4,  \cite{Kal04,Kal05,Loo21b,Loo22} etc. However, 
the original QUEST database presents inherent limitations, focusing on transition energies, and featuring a restricted range of small organic compounds.  Over the last three years, our research efforts have been dedicated to expanding 
QUEST in various directions. This includes incorporating EES with a substantial charge-transfer (CT) character, \cite{Loo21a} exploring compounds with an inverted singlet-triplet gap, \cite{Loo23a} defining ultra-accurate EES dipole 
moments and oscillator strengths for small compounds, \cite{Chr21,Sar21,Dam22} treating systems containing transition metals, \cite{Loo23b} and introducing bicyclic organic derivatives. \cite{Loo21}

\begin{figure}[htp]
  \includegraphics[width=\linewidth]{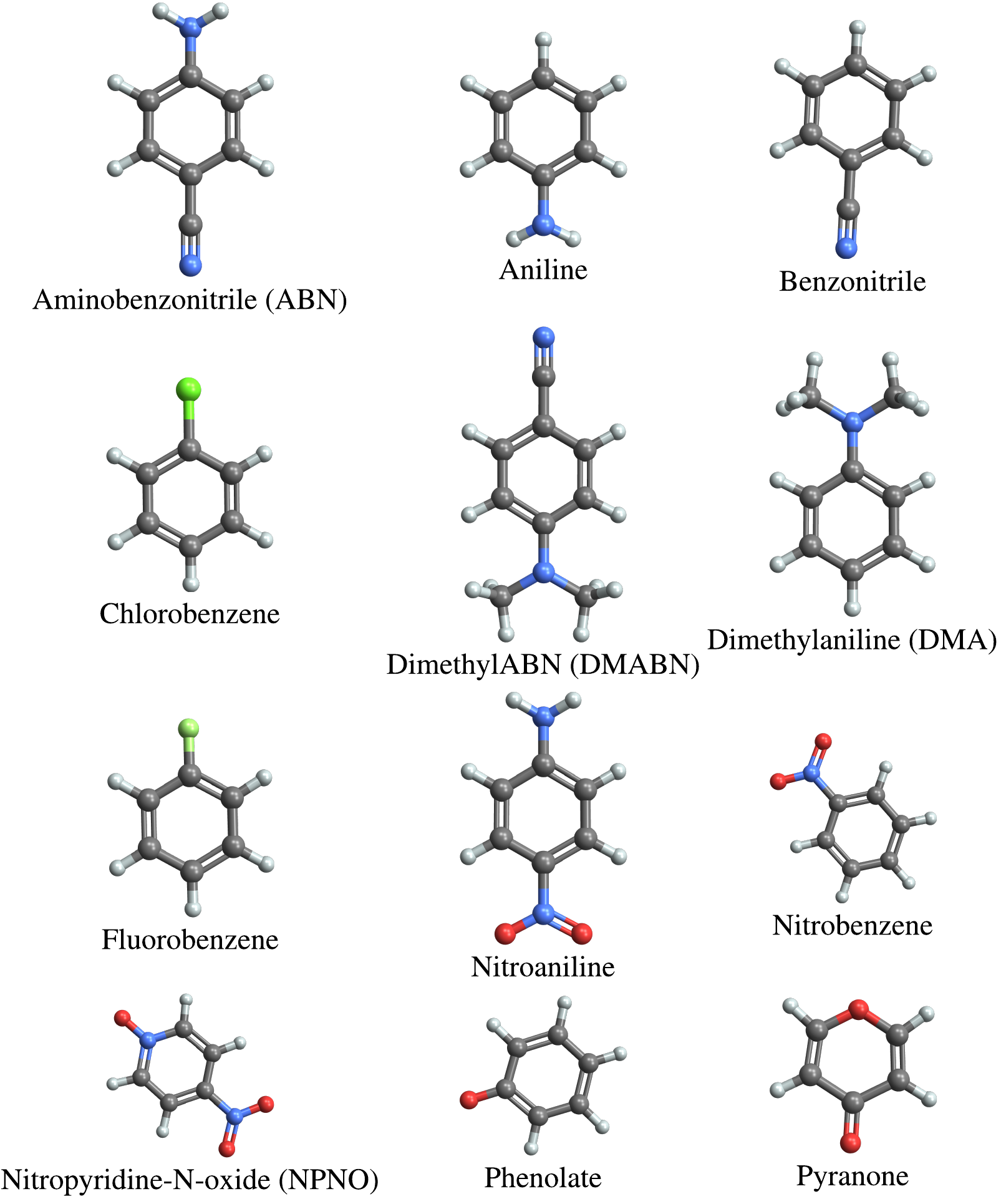}
  \caption{Representation of the systems investigated in the present study.}
  \label{Fig-1}
\end{figure}

In the present contribution, we explore a previously overlooked aspect: the impact of chemical substitution on a given chemical core. Indeed, when using efficient theoretical methods, such as TD-DFT or CC2, for solving chemical
problems, trends within a homologous series often become the central question. In this framework, we focus here on the most popular conjugated unit, the phenyl ring, examining various substitution patterns (mono- and di-substituted, 
with donor and/or acceptor groups), as can be seen in Figure \ref{Fig-1}.  For all EES, we obtained, at the very least, CC3/{\AVTZ} estimates, often corrected thanks to CCSDT double-$\zeta$ results, and even CC4 contributions
in a few cases. Except for the CT transitions in a few compounds, \cite{Gui18b,Loo21a} these derivatives have, to the very best of our knowledge, never been treated at these high levels of theory.

This paper is organized as follows. First, we describe the theoretical protocols employed to define the theoretical best estimates (TBEs) associated with these vertical excitation energies and the software used in the benchmarking section. 
Next, we compare these TBEs with literature data and we discuss their accuracy. Next, we evaluate the performances of lower-order models using these TBEs and consider both absolute and relative accuracies.  Finally, we summarize 
our main findings.

\section{Computational details}
\label{sec:Comp}

\subsection{Geometries}

The ground-state geometry of all systems has been optimized, depending on the size of the compound, at the CCSD(T) or CC3 level using the cc-pVTZ atomic basis set and applying the frozen-core (FC) approximation. These optimizations have 
been achieved with CFOUR,  \cite{cfour,Mat20} relying on default convergence thresholds. Cartesian coordinates for all structures are available in the Supporting Information (SI). It is noteworthy that for obvious computational reasons, 
we enforce symmetry constraints, i.e.,  all compounds of Figure \ref{Fig-1} have been considered in the $C_{2v}$ point group. While this is the correct point group in most cases, we underline that the amino group can be slightly puckered in some systems.
For example, aniline should normally be of $C_s$ symmetry as the nitrogen lone pair is not fully delocalized on the phenyl cycle. 

\subsection{Definition of the TBE}

To determine the TBEs associated with the VTEs, we relied on high-level CC calculations, all performed in the FC approximation. We carried out the CC calculations using either the equation-of-motion (EOM) or linear-response (LR) formalisms, but we do 
not specify this aspect below since both yield the very same transition energies.  \cite{Row68,Sta93} We performed CC3, CCSDT, and CC4 calculations, using three atomic basis sets containing both polarization and diffuse functions, namely 
{\Pop}, {\AVDZ}, and {\AVTZ},  which are simply denoted ``Pop'',  ``AVDZ'', and ``AVTZ'' below.  To perform these CC calculations we used CFOUR, \cite{cfour,Mat20}  DALTON, \cite{dalton} and MRCC. \cite{Kal20,mrcc} The former allows for CC4
calculations but only treats singlet excited states. The triplet transition energies have been obtained with Dalton (CC3) and MRCC (CCSDT). For the cases allowing for direct comparisons between the three codes, we have not found VTE variations 
exceeding 0.001 eV when applying default convergence thresholds for the three software packages.

In most cases, we could achieve CCSDT calculations with one of the double-$\zeta$ basis sets, hence our TBEs were obtained as
\begin{equation}
\begin{split}
	\label{eq-t-pop}
	\Delta {E}_{\text{AVTZ}}^{\text{TBE}} & =
	\Delta \tilde{E}_{\text{AVTZ}}^{\text{CCSDT}} 
	\\
	& \simeq \Delta E_{\text{AVTZ}}^{\text{CC3}} + \left[ \Delta E_{\text{Pop}}^{\text{CCSDT}} - \Delta E_{\text{Pop}}^{\text{CC3}} \right],
\end{split}
\end{equation}
or
\begin{equation}
\begin{split}
	\label{eq-t-dz}
	\Delta {E}_{\text{AVTZ}}^{\text{TBE}} & = 
	\Delta \tilde{E}_{\text{AVTZ}}^{\text{CCSDT}} 
	\\
	& \simeq \Delta E_{\text{AVTZ}}^{\text{CC3}} + \left[ \Delta E_{\text{AVDZ}}^{\text{CCSDT}} - \Delta E_{\text{AVDZ}}^{\text{CC3}} \right].
\end{split}
\end{equation}
These equations have been demonstrated to provide excellent estimates of the actual CCSDT/{\AVTZ} values. \cite{Loo22} However, EES with a significant CT (sCT) character deserve a specific treatment. Indeed, it has been shown, first for 
intermolecular \cite{Koz20} and next for intramolecular \cite{Loo21a} transitions having a sCT nature that CCSDT-3 \cite{Wat96,Pro10} often delivers VTEs closer to CCSDT than CC3, which is an unusual outcome. \cite{Ver21} Therefore for 
these transitions, we apply the same two equations as above, but using CCSDT-3 rather than CC3 VTEs in the right-hand side. 

For some EES of the most compact molecules, we could achieve CC4/6-31+G(d) calculations for the singlet EES, and we therefore computed the TBEs as
\begin{equation}\label{eq-4}
\begin{split}
	\Delta {E}_{\text{AVTZ}}^{\text{TBE}}  
	& = \Delta \tilde{E}_{\text{AVTZ}}^{\text{CC4}} 
	\\
	& \simeq \Delta E_{\text{AVTZ}}^{\text{CC3}} + \left[ \Delta E_{\text{AVDZ}}^{\text{CCSDT}} - \Delta E_{\text{AVDZ}}^{\text{CC3}}\right] 
	\\
	&+   \left[ \Delta E_{\text{Pop}}^{\text{CC4}} - \Delta E_{\text{Pop}}^{\text{CCSDT}} \right].
\end{split}
\end{equation}
We underline that CC4 provides results very close to both CCSDTQ and FCI, \cite{Loo21b,Loo22} so that the VTE defined with Eq.~\eqref{eq-4} can be considered chemically accurate.

Finally, for the triplet EES of the largest compounds, we have been computationally limited to CC3/aug-cc-pVTZ. Hence, we simply relied on the following definition:
\begin{equation}
	\label{eq-3}
	\Delta {E}_{\text{AVTZ}}^{\text{TBE}}  = 
	\Delta {E}_{\text{AVTZ}}^{\text{CC3}}.
\end{equation}
Note that triplet EES are typically characterized by a very large single excitation character ($\%T_1$),  \cite{Sch08,Ver21} so one can expect CC3 to provide a very satisfying description for these states.

\subsection{Benchmarks}

We have used the TBEs defined in the previous Section to benchmark a large set of wave function methods. Again, we systematically applied the FC approximation and used the aug-cc-pVTZ atomic basis set. 

We performed CIS(D), \cite{Hea94,Hea95} CC2,  \cite{Chr95,Hat00} SOS-ADC(2), SOS-CC2, and SCS-CC2  \cite{Hel08} calculations with  Turbomole 7.3, \cite{Turbomole,Bal20} using the resolution-of-the-identity (RI) scheme with the RI-{\AVTZ}  
auxiliary basis set. \cite{Wei02} Default scaling parameters were used for the SOS and SCS calculations. The EOM-MP2,  \cite{Sta95c}   ADC(2), \cite{Tro97,Dre15} SOS-ADC(2) \cite{Kra13} and ADC(3) \cite{Tro02,Har14,Dre15} VTEs have been 
obtained with Q-CHEM 6.0, \cite{Epi21} applying the RI approximation with the corresponding auxiliary basis set, and setting the threshold for integral accuracy to $10^{-14}$. The SOS parameters differ in Turbomole and Q-Chem, hence the 
corresponding SOS-ADC(2) results are respectively tagged  [TM] and [QC] in the following. Having the ADC(2) and ADC(3) values at hand, the ADC(2.5) results are simply obtained as their average, \cite{Loo20b} since the optimal mixing
between the two approaches is close to the 50/50 ratio. \cite{Bau22} Similarity-transformed EOM-CCSD (STEOM-CCSD) \cite{Noo97,Dut18}  energies were computed with ORCA 5.0. \cite{Nee20}  We checked that the active-space percentage was, 
at least,  $98\%$ for all reported EES. CCSD \cite{Pur82}  calculations were achieved with Gaussian 16, \cite{Gaussian16} whereas we used Dalton \cite{dalton} to obtain the CCSDR(3) VTEs.  \cite{Chr96b} Finally, CFOUR, \cite{cfour,Mat20}
was employed to access both CCSD(T)(a)$^\star$ \cite{Mat16} and CCSDT-3 \cite{Wat96,Pro10} VTEs.

\section{Results and Discussion}
\label{sec:Res}

\subsection{Reference values}

Our TBEs are listed in Tables \ref{Table-1} and \ref{Table-2}, where we also report some key features for each EES: the oscillator strength ($f$) and $\%T_1$ as obtained at CC3/aug-cc-pVDZ level with Dalton, and the size change of the electronic cloud 
estimated at the ADC(3)/aug-cc-pVTZ level with Q-Chem.  Further details, such as key molecular orbitals, and raw CC3/CCSDT/CC4 values are available in the SI. The data of Tables \ref{Table-1} and \ref{Table-2} represent a total of 108 VTEs, 
including 72 singlet and 36 triplet transitions, 74 valence (58 $\ppi$ and 16 $\npi$, 7 showing a significant CT character) and 32 Rydberg EES, 2 states showing a strongly mixed nature.

\begin{squeezetable}
\begin{table*}[htp]
\caption{Theoretical best estimates (TBEs) of the VTEs (in eV) for ABN, aniline, benzonitrile, chlorobenzene, DMABN, and DMA. The extrapolation procedure employed to determine the TBE, the percentage of single excitation ($\%T_1$) 
computed at the CC3/aug-cc-pVTZ (in \%), and oscillator strength ($f$) are given, together with the variation of the size of the electronic cloud during the excitation process ($\Delta r^2$ in a.u.) computed at ADC(3)/aug-cc-pVTZ level. 
Note that the $f$ have been obtained using the LR-CC3 formalism. A selection of literature data is given in the right-most column. The acronyms sCT, wCT, and Ryd stand
for significant CT, weak CT, and Rydberg, respectively.}
\label{Table-1}
\begin{ruledtabular}
\begin{tabular}{llclcccl}
Compound	& State				&\mc{2}{c}{TBE}				&$\%T_1$	&$f$	&$\Delta r^2$	&Literature	\\
\hline	
ABN			&$^1B_2$ ($\ppi$)		&4.448	& Eq.~\eqref{eq-t-pop}	&86.8	&0.018	&-1					&4.26 [CASPT2]\cite{Sob96}; 4.27 [STEOM] \cite{Par99}; $\sim$4.27 [$\lmax^\text{heptane}$] \cite{Zac93}\\
			&$^1B_1$ (Ryd)		&5.032	& Eq.~\eqref{eq-t-pop}	&92.3	&0.002	&5					&\\
			&$^1A_1$ (wCT, $\ppi$)	&5.092	& Eq.~\eqref{eq-t-pop}	&89.2	&0.427	&3					&4.98 [CASPT2] \cite{Sob96}; 5.13 [STEOM] \cite{Par99}; 4.73 [$\lmax^\text{heptane}$] \cite{Zac93}\\
			&$^1A_2$ (Ryd)		&5.783	& Eq.~\eqref{eq-t-pop}	&92.4	&		&61					&\\
			&$^1B_1$ (Ryd)		&5.827	& Eq.~\eqref{eq-t-pop}	&92.1	&		&45					&\\
			&$^3A_1$ ($\ppi$)		&3.635	& Eq.~\eqref{eq-3}		&97.6	&		&0					&3.25 [STEOM] \cite{Par99}\\
			&$^3B_2$ ($\ppi$)		&4.126	&Eq.~\eqref{eq-3}		&97.0	&		&-3					&4.11 [STEOM] \cite{Par99}\\
Aniline		&$^1B_2$ ($\ppi$)		&4.504	& Eq.~\eqref{eq-t-dz}	&87.3	&0.029	&0					&4.31 [CCSD(T)] \cite{Wan13e};  4.33 [CASPT2] \cite{Hou05}; 4.40  [$\lmax^\text{gas}$]\cite{Kim64}; 4.22 [$\eoo^\text{gas}$] \cite{Eba02}\\
			&$^1B_1$ (Ryd)		&4.721	& Eq.~\eqref{eq-t-dz}	&92.6	&0.004	&46					&4.69 [CCSD(T)] \cite{Wan13e};  4.85[CASPT2] \cite{Hou05}; 4.53 [SACCI]; \cite{Hon02b}4.60 [$\eoo^\text{gas}$] \cite{Eba02}\\
			&$^1B_1$ (Ryd)		&5.415	& Eq.~\eqref{eq-t-dz}	&92.4	&0.000	&58					&5.31 [CCSD(T)] \cite{Wan13e}; 5.99[CASPT2] \cite{Hou05} \\
			&$^1A_2$ (Ryd)		&5.437	& Eq.~\eqref{eq-t-dz}	&92.7	&		&60					&5.42 [CCSD(T)] \cite{Wan13e}; 5.92[CASPT2] \cite{Hou05} \\
			&$^1A_1$ (Mixed)		&5.501	& Eq.~\eqref{eq-t-dz}	&91.0	&0.177	&7					&5.42 [CCSD(T)] \cite{Wan13e}; 5.54 [CASPT2] \cite{Hou05}; 5.34 [SACCI] \cite{Hon02b}; 5.39  [$\lmax^\text{gas}$]\cite{Kim64}  \\
			&$^3A_1$ ($\ppi$)		&3.951	& Eq.~\eqref{eq-t-pop}	&98.1	&		&1					&3.74 [CASPT2] \cite{Hou05}\\
			&$^3B_2$ ($\ppi$)		&4.108	& Eq.~\eqref{eq-t-pop}	&97.3	&		&0					&3.82 [CASPT2] \cite{Hou05}\\
			&$^3B_1$ (Ryd)		&4.637	& Eq.~\eqref{eq-t-pop}	&97.4	&		&45					&4.60 [CASPT2] \cite{Hou05}\\
			&$^3A_1$ ($\ppi$)		&4.653	& Eq.~\eqref{eq-t-pop}	&97.3	&		&1					&4.40 [CASPT2] \cite{Hou05}\\
			&$^3B_1$ (Ryd)		&5.387	& Eq.~\eqref{eq-t-pop}	&97.3	&		&55					&\\
Benzonitrile	&$^1B_2$ ($\ppi$)		&4.864	& Eq.~\eqref{eq-t-dz}	&85.8	&0.003	&2					&4.64 [CASPT2] \cite{Sob96}; 5.01 [CCSD] \cite{Med15}; 4.53 [$\eoo^\text{gas}$] \cite{Raj22}\\
			&$^1A_1$ ($\ppi$)		&6.026	& Eq.~\eqref{eq-t-dz}	&91.9	&0.214	&3					&6.24 [CASPT2] \cite{Sob96}; 6.12 [CCSD] \cite{Med15}; 5.53 [$\eoo^\text{gas}$] \cite{Raj22}\\
			&$^1B_1$ (Ryd)		&6.833	& Eq.~\eqref{eq-t-dz}	&92.5	&0.006	&42					&5.57 [$\eoo^\text{gas}$] \cite{Raj22}\\
			&$^1B_2$ ($\ppi$)		&6.902	& Eq.~\eqref{eq-t-dz}	&90.8	&0.426	&5					&6.53 [$\lmax^\text{gas}$] \cite{Raj22}\\
			&$^1A_2$ (Ryd)		&6.979	& Eq.~\eqref{eq-t-dz}	&92.6	&		&45					&\\
			&$^1A_1$ ($\ppi$)		&6.979	& Eq.~\eqref{eq-t-dz}	&91.2	&0.604	&3					&6.62 [$\lmax^\text{gas}$] \cite{Raj22}\\
			&$^1A_2$ (wCT, $\ppi$)	&7.052	& Eq.~\eqref{eq-t-dz}	&90.5	&		&-2					&7.37 [CASPT2] \cite{Sob96}\\
			&$^3A_1$ ($\ppi$)		&3.888	& Eq.~\eqref{eq-t-pop}	&98.2	&		&1					&3.86 [CC3] \cite{Med15} \\
			&$^3B_2$ ($\ppi$)		&4.654	& Eq.~\eqref{eq-t-pop}	&96.9	&		&3					&4.65 [CC3] \cite{Med15} \\
			&$^3A_1$ ($\ppi$)		&4.702	& Eq.~\eqref{eq-t-pop}	&97.1	&		&2					&\\
			&$^3B_2$ ($\ppi$)		&5.654	& Eq.~\eqref{eq-t-pop}	&97.8	&		&2					&\\		
Chlorobenzene	&$^1B_2$ ($\ppi$)		&4.929	& Eq.~\eqref{eq-t-dz}	&86.3 	&0.001	&1					&4.64 [CASPT2]\cite{Liu04b}; 4.59 [$\eoo^\text{gas}$]\cite{Pal16b}; 4.71  [$\lmax^\text{gas}$]\cite{Kim65} \\	
			&$^1A_1$ ($\ppi$)		&6.197	& Eq.~\eqref{eq-t-dz}	&92.3	&0.081	&4					&6.26 [CASPT2]\cite{Liu04b};  5.95  [$\lmax^\text{gas}$]\cite{Kim65}\\	
			&$^1B_1$ (Ryd)		&6.316	& Eq.~\eqref{eq-t-dz}	&92.4	&0.081	&38					&6.46 [CASPT2]\cite{Liu04b}\\	
			&$^1B_1$ (mixed)		&6.717	& Eq.~\eqref{eq-t-dz}	&91.9	&0.004	&25					&\\	
			&$^1A_2$ (Ryd)		&6.767	& Eq.~\eqref{eq-t-dz}	&92.4	&		&45					&\\	
			&$^1B_2$ ($\ppi$)		&6.902	& Eq.~\eqref{eq-t-dz}	&91.0	&0.637	&6					&6.83 [CASPT2]\cite{Liu04b}; 6.72  [$\lmax^\text{gas}$]\cite{Kim65}\\	
			&$^1A_1$ ($\ppi$)		&7.011	& Eq.~\eqref{eq-t-dz}	&91.6	&0.637	&5					& 6.88  [$\lmax^\text{gas}$]\cite{Kim65}\\	
			&$^1A_2$ (Ryd)		&7.018	& Eq.~\eqref{eq-t-dz}	&92.6	&		&65					&\\	
			&$^3A_1$ ($\ppi$)		&4.073	& Eq.~\eqref{eq-t-pop}	&98.4	&		&1					&3.58 [CASPT2]\cite{Liu04b}\\	
			&$^3B_2$ ($\ppi$)		&4.723	& Eq.~\eqref{eq-t-pop}	&97.0	&		&1					&4.48 [CASPT2]\cite{Liu04b}\\	
			&$^3A_1$ ($\ppi$)		&4.768	& Eq.~\eqref{eq-t-pop}	&97.0	&		&1					&4.31 [CASPT2]\cite{Liu04b}\\	
			&$^3B_2$ ($\ppi$)		&5.725	& Eq.~\eqref{eq-t-pop}	&97.8	&		&3					&5.61 [CASPT2]\cite{Liu04b}\\	
DMABN		&$^1B_2$ (Val, $\ppi$)	&4.336	& Eq.~\eqref{eq-t-pop}	&		&		&-2					&4.32 [MRCI]\cite{Geo15}; 4.47 [ADC(3)]\cite{Mew17}; 4.48 [CCSD(T)(a)]:\cite{Kny22} 4.00 [$\eoo^\text{gas}$]\cite{Dru10} \\
			&$^1B_1$ (Ryd)		&4.816	& Eq.~\eqref{eq-t-pop}	&		&		&43					&\\
			&$^1A_1$ (sCT, $\ppi$)	&4.866	& Eq.~\eqref{eq-t-pop}\footnotemark[1]	&		&		&3				&4.90 [MRCI]\cite{Geo15}; 4.94 [ADC(3)]\cite{Mew17}; 5.00 [CCSD(T)(a)]:\cite{Kny22} 4.57 [$\lmax^\text{gas}$]\cite{Dru10} \\	
			&$^1A_2$ (Ryd)		&5.457	& Eq.~\eqref{eq-t-pop}	&		&		&64					&\\
DMA			&$^1B_2$ ($\ppi$)		&4.396	& Eq.~\eqref{eq-t-pop}	&87.2	&		&2					&4.65 [CASPT2]\cite{Gal09}; 4.49 [CCSDR(3)]\cite{Tho15}; 4.30 [$\lmax^\text{gas}$]\cite{Kim64}\\	
			&$^1B_1$ (Ryd)		&4.517	& Eq.~\eqref{eq-t-pop}	&92.1	&		&45					&4.54 [CCSDR(3)]\cite{Tho15} \\	
			&$^1A_2$ (Ryd)		&5.114	& Eq.~\eqref{eq-t-pop}	&92.0	&		&68					&5.13 [CCSDR(3)]\cite{Tho15} \\	
			&$^1B_1$ (Ryd)		&5.160	& Eq.~\eqref{eq-t-pop}	&		&		&80					&5.16 [CCSDR(3)]\cite{Tho15} \\	
			&$^1A_1$ (Ryd)		&5.333	& Eq.~\eqref{eq-t-pop}	&92.3	&		&49					&\\	
			&$^1A_1$ ($\ppi$)		&5.402	& Eq.~\eqref{eq-t-pop}	&		&		&16					&5.32 [CASPT2]\cite{Gal09}; 5.15 [$\lmax^\text{gas}$]\cite{Kim64}\\\	
			&$^3A_1$ ($\ppi$)		&3.970	&Eq.~\eqref{eq-3}		&98.0	&		&1					&\\	
			&$^3B_2$ ($\ppi$)		&4.000	&Eq.~\eqref{eq-3}		&97.1	&		&2					&\\	
\end{tabular}
\end{ruledtabular}
\footnotetext[1]{TBE determined via Eq.~\eqref{eq-t-pop} using CCSDT-3 data (see main text for more details).}
\end{table*}
\end{squeezetable}

\begin{squeezetable}
\begin{table*}[htp]
\caption{Theoretical best estimates (TBEs) of the VTEs (in eV) for fluorobenzene, nitroaniline, nitrobenzene, NPNO, phenolate, and pyranone.  See caption of Table \ref{Table-1} for more details.}
\label{Table-2}
\begin{ruledtabular}
\begin{tabular}{llclcccl}
Compound	& State				&\mc{2}{c}{TBE}		&$\%T_1$	&$f$	&$\Delta r^2$	&Literature	\\
\hline	
Fluorobenzene	&$^1B_2$ ($\ppi$)		&5.006	& Eq.~\eqref{eq-4}		&86.4	&0.007	&2					&4.74 [CASPT2]\cite{Liu04b}; 4.69 [$\eoo^\text{gas}$] \cite{Phi81}; 4.84  [$\lmax^\text{gas}$]\cite{Kim65} \\	
			&$^1A_1$ ($\ppi$)		&6.426	& Eq.~\eqref{eq-4}		&92.5	&0.000	&4					&6.40 [CASPT2]\cite{Liu04b}; 6.21  [$\lmax^\text{gas}$]\cite{Phi81}; 6.19  [$\lmax^\text{gas}$]\cite{Kim65} \\	
			&$^1B_1$ (Ryd)		&6.475	& Eq.~\eqref{eq-4}		&92.8	&0.003	&47					&6.32[$\eoo^\text{gas}$]\cite{Pal16} \\	
			&$^1A_2$ (Ryd)		&6.831	& Eq.~\eqref{eq-4}		&92.8	&		&47					&\\	
			&$^1B_1$ (Ryd)		&7.057	& Eq.~\eqref{eq-4}		&93.1	&0.022	&54					&6.89[$\eoo^\text{gas}$]\cite{Pal16} \\				
			&$^1A_2$ (Ryd)		&7.095	& Eq.~\eqref{eq-t-dz}	&92.6	&		&65					&\\
			&$^1B_2$ ($\ppi$)		&7.192	& Eq.~\eqref{eq-t-dz}	&91.2	&0.556	&8					& \\
			&$^1A_1$ ($\ppi$)		&7.225	& Eq.~\eqref{eq-t-dz}	&91.6	&0.584	&6					&7.50 [CCSD]\cite{Miu07}; 6.99  [$\lmax^\text{gas}$]\cite{Phi81}; 7.00  [$\lmax^\text{gas}$]\cite{Kim65}\\
			&$^3A_1$ ($\ppi$)		&4.184	& Eq.~\eqref{eq-t-pop}	&98.5	&		&1					&3.75 [CASPT2]\cite{Liu04b} \\
			&$^3B_2$ ($\ppi$)		&4.750	& Eq.~\eqref{eq-t-pop}	&97.1	&		&1					&4.53 [CASPT2]\cite{Liu04b} \\
			&$^3A_1$ ($\ppi$)		&4.869	& Eq.~\eqref{eq-t-pop}	&97.0	&		&1					&4.52 [CASPT2]\cite{Liu04b} \\
			&$^3B_2$ ($\ppi$)		&5.932	& Eq.~\eqref{eq-t-pop}	&97.9	&		&3					&5.92 [CASPT2]\cite{Liu04b} \\
Nitroaniline	&$^1A_2$ ($\npi$)		&3.998	&Eq.~\eqref{eq-t-pop}	&87.5	&		&-3					&\\
			&$^1A_1$ (sCT, $\ppi$)	&4.402	&Eq.~\eqref{eq-t-pop}\footnotemark[1]	&87.2	&		&3				&4.54 [CCSDR(3)]\cite{Cas19}; 4.30 [CC(2,3)]\cite{Hoy16}; 3.72 [CASPT2]\cite{San22b}; 3.59 [$\lmax^\text{benzene}$]\cite{Sch96b}\\
			&$^1B_2$ ($\ppi$)		&4.508	&Eq.~\eqref{eq-t-pop}	&86.7	&		&0					&4.46 [CASPT2]\cite{San22b}\\
			&$^1B_1$  ($\npi$)		&4.507	&Eq.~\eqref{eq-t-pop}	&87.0	&		&-3					&\\
			&$^3A_1$ ($\ppi$)		&3.445	&Eq.~\eqref{eq-3}		&97.2	&		&2					&3.41 [CCSD]\cite{Nan16}\\
			&$^3A_2$  ($\npi$)		&3.793	&Eq.~\eqref{eq-3}		&96.5	&		&-3					&\\
			&$^3B_2$ ($\ppi$)		&3.798	&Eq.~\eqref{eq-3}		&98.0	&		&-3					&\\	
Nitrobenzene	&$^1A_2$ ($\npi$)		&3.926	& Eq.~\eqref{eq-t-dz}	&87.4	&		&-1					&3.32 [CASPT2]\cite{Giu17};  3.94 [CASPT2]\cite{Sot21}; 4.00 [ADC(3)]\cite{Mew14}; 3.65 [$\lmax^\text{gas}$]\cite{Nag64}\\
			&$^1B_1$ ($\npi$)		&4.390	& Eq.~\eqref{eq-t-dz}	&87.0	&0.000	&-1					&3.84 [CASPT2]\cite{Giu17}; 4.37  [CASPT2]\cite{Sot21}; 4.40 [ADC(3)]\cite{Mew14}\\
			&$^1B_2$ ($\ppi$)		&4.725	& Eq.~\eqref{eq-t-dz}	&86.1	&0.007	&3					&4.37 [CASPT2]\cite{Giu17}; 4.79 [CASPT2]\cite{Sot21}; 4.67 [ADC(3)]\cite{Mew14}\\
			&$^1A_1$ (sCT, $\ppi$)	&5.410	& Eq.~\eqref{eq-t-dz}\footnotemark[2]	&88.7	&0.220	&4				&5.01 [CASPT2]\cite{Giu17};  5.22 [CASPT2]\cite{Sot21};  5.27 [ADC(3)]\cite{Mew14}; 5.15 [$\lmax^\text{gas}$]\cite{Kri16}\\
			&$^3B_2$ ($\ppi$)		&3.625	&Eq.~\eqref{eq-t-pop}	&98.0	&		&-1					&3.30 [CASPT2]\cite{Giu17}; 3.17 [ADC(3)]\cite{Mew14}\\
			&$^3A_2$ ($\npi$)		&3.685	&Eq.~\eqref{eq-t-pop}	&96.6	&		&-1					&3.15 [CASPT2]\cite{Giu17}; 3.69 [ADC(3)]\cite{Mew14}\\
			&$^3A_1$ ($\ppi$)		&3.915	&Eq.~\eqref{eq-t-pop}	&98.0	&		&3					&3.71 [CASPT2]\cite{Giu17}; 3.64 [ADC(3)]\cite{Mew14}\\
			&$^3B_1$ ($\npi$)		&4.204	&Eq.~\eqref{eq-t-pop}	&96.6	&		&-1					&3.75 [CASPT2]\cite{Giu17}\\
NPNO		&$^1A_2$ ($\npi$)		&3.794	&Eq.~\eqref{eq-t-pop}	&83.4	&		&-9					&\\
			&$^1A_2$ ($\npi$)		&3.978	&Eq.~\eqref{eq-t-pop}	&85.7	&		&-2					&\\
			&$^1B_2$ ($\ppi$)		&4.006	&Eq.~\eqref{eq-t-pop}	&81.7	&		&0					&\\
			&$^1A_1$ (sCT,$\ppi$)	&4.078	&Eq.~\eqref{eq-t-pop}\footnotemark[1]&84.4	&		&1					&4.32 [CASPT2]\cite{Bud16b};  3.80 [$\lmax^\text{gas}$]\cite{Lag96,Bud16b} \\
			&$^1B_1$ ($\npi$)		&4.395	&Eq.~\eqref{eq-t-pop}	&85.7	&		&-2					&\\
			&$^1B_1$ ($\npi$)		&5.138	&Eq.~\eqref{eq-t-pop}	&78.5	&		&-13					&\\
			&$^3A_1$ (wCT, $\ppi$)	&2.469	&Eq.~\eqref{eq-3}		&97.1	&		&-1					&\\
			&$^3A_2$ ($\npi$)		&3.634	&Eq.~\eqref{eq-3}		&95.3	&		&-2					&\\
			&$^3B_2$ ($\ppi$)		&3.693	&Eq.~\eqref{eq-3}		&97.8	&		&-2					&\\
			&$^3A_2$ ($\npi$)		&3.804	&Eq.~\eqref{eq-3}		&95.5	&		&-9					&\\
			&$^3B_2$ ($\ppi$)		&3.876	&Eq.~\eqref{eq-3}		&95.2	&		&-2					&\\
Phenolate		&$^1B_1$ (Ryd)		&2.610	& Eq.~\eqref{eq-4}		&90.8	&0.001	&80					&\\
			&$^1A_2$ (Ryd)		&2.903	& Eq.~\eqref{eq-4}		&90.9	&		&99		 			&\\
			&$^1B_1$ (Ryd)		&3.064	& Eq.~\eqref{eq-4}		&91.2	&0.005	&99					&\\
			&$^1A_2$ (Ryd)		&3.437	& Eq.~\eqref{eq-t-dz}	&91.3	&		&115					&\\
			&$^1B_1$ (Ryd)		&3.486	& Eq.~\eqref{eq-t-dz}	&91.2	&0.001	&122					&\\
			&$^1B_2$ (Ryd)		&3.759	& Eq.~\eqref{eq-t-dz}	&85.0	&0.000	&69					&\\
			&$^1B_2$  ($\ppi$)		&3.853	& Eq.~\eqref{eq-t-dz}	&87.3	&0.050	&6					&\\
			&$^1A_1$ (Ryd)		&3.967	& Eq.~\eqref{eq-t-dz}	&85.5	&0.010	&95					&\\
			&$^1A_1$  ($\ppi$)		&4.205	& Eq.~\eqref{eq-t-dz}	&90.8	&0.009	&31					&\\
Pyranone		&$^1A_2$ ($\npi$)		&4.009	& Eq.~\eqref{eq-4}		&86.3	&		&-4					& 3.52 [$\eoo^\text{gas}$]\cite{Ses19}\\
			&$^1B_1$ ($\npi$)		&5.178	& Eq.~\eqref{eq-4}		&81.7	&0.000	&-1					&\\
			&$^1A_1$ ($\ppi$)		&5.617	& Eq.~\eqref{eq-4}		&86.6	&0.234	&0					&\\
			&$^1B_2$ ($\ppi$)		&5.738	& Eq.~\eqref{eq-4}		&87.1	&0.015	&-1					&\\
			&$^1B_2$ (Ryd)		&6.459	& Eq.~\eqref{eq-4}		&86.2	&0.006	&38					&\\
			&$^1B_1$ (Ryd)		&6.539	& Eq.~\eqref{eq-4}		&91.1	&0.027	&38					&\\
			&$^3A_2$ ($\npi$)		&3.819	& Eq.~\eqref{eq-t-pop}	&96.3	&		&-4					& 3.38 [$\eoo^\text{gas}$]\cite{Par23}\\
			&$^3A_1$ ($\ppi$)		&3.967	& Eq.~\eqref{eq-t-pop}	&97.9	&		&-2					&\\
			&$^3B_2$ ($\ppi$)		&4.431	& Eq.~\eqref{eq-t-pop}	&98.0	&		&-1					&\\
\end{tabular}
\end{ruledtabular}
\footnotetext[1]{TBE determined via Eq.~\eqref{eq-t-pop} using CCSDT-3 data (see main text for more details).}
\footnotetext[2]{TBE determined via Eq.~\eqref{eq-t-dz} using CCSDT-3 data (see main text for more details).}
\end{table*}
\end{squeezetable}

In the right-most column of Tables \ref{Table-1} and \ref{Table-2}, we provide previous theoretical and experimental estimates for comparison. Caution is advised when comparing with previous computational data, given the different ground-state 
geometries and basis sets. This caution extends to measured values, whether given as 0-0 energies or wavelengths of maximal absorption ($\lmax$), as these do not correspond to vertical transition energies (see discussion in Section \ref{sec:Intro}). 
Nonetheless, one can anticipate qualitative consistency in chemical trends, excited-state ordering, and absolute energies.

For the amino-substituted benzonitriles, that is, ABN, \cite{Ser95,Sob96,Ser97b,Par99,Gom15,Seg16,Seg16b,Cas18} and DMABN,
\cite{Ser95,Sob96,Roo96,Roo97,Ser97b,Par99,Pea08,Hel08,Fde09,Wig09,Rhe09,Ngu10,Fde10,Ngu11,Mar11,Pea12,Dev12,Hed13,Lun13,Geo15,Gom15,Hoy16,Seg16,Seg16b,Mew17,Jac17b,Gui18b,Mew18,Cas19,Car19,Kny22}
previous theoretical studies focused on the competition between local and charge-transfer EES in the emission process, or the magnitude of the charge transfer in the latter state. The present TBEs for these two key $\ppi$ transitions 
agree well with the most accurate previous studies, although one exception arises with the lowest triplet, appearing too low with STEOM-CCSD. \cite{Par99} Rydberg EES were not reported in previous works due to the use of diffuse-less basis sets.

For aniline, comprehensive sets of VTEs have been reported, \cite{Hou05,Wan13e,Tho15} and the present TBEs align well with the latter study as can be seen in Table \ref{Table-1}. Additional  \emph{ab initio} investigations of EES for 
aniline are also available.  \cite{Hon02b,Miu07,Sal14,Rob21}

Regarding the singlet valence transitions of benzonitrile, the present data are reasonably close to previous CASPT2 \cite{Sob96} and CC \cite{Med15} estimates. A perfect match is logically observed with the CC3 data of Ref.~\citenum{Med15} 
for the corresponding triplet excitations. For other EES, where only experimental data are available, the present VTEs are consistently larger than all measured 0-0 energies in the gas phase, \cite{Raj22} with a particularly substantial correction
for Rydberg transitions, which is sound.

For the two monohalobenzenes, the previous extensive CASPT2 investigation of Liu and co-workers \cite{Liu04b} provides similar trends as the current study, though the CASPT2 energies for the three lowest triplet EES were notably 
underestimated in this earlier work. Detailed joint theoretical-experimental studies \cite{Pal16,Pal16b} and other \emph{ab initio} works \cite{Miu07,Rob21} are available for a few specific EES.

For dimethylaniline, previous wave function studies using CASPT2 \cite{Gal09} and CCSDR(3) \cite{Tho15} deliver the same EES ordering as the present study. Notably, the use of diffuse functions here slightly obscures the CT
character detected previously in the $^1A_1$ EES, \cite{Loo21a} and we therefore characterized this state as $\ppi$ in Table \ref{Table-1}.

Previous works devoted to the push-pull nitroaniline mostly  focused on the CT EES, \cite{Ser97b,Hoy16,Nan16,Lu18,Gui18b,Cas19,Fol20,Loo21a,San22b} and our TBE agrees well with previous CC results, while the recent 
XMS-CASPT2 value of 3.72 eV for this $\ppi$ state \cite{San22b} seems too low. More comprehensive data can be found for the parent nitrobenzene \cite{Kro00,And08,Que11,Mew14,Kri16,Giu17,Sch18b,Sot21} and it can be seen in Table \ref{Table-2}
that the current TBEs globally fit the trends obtained earlier, though some significant changes in absolute VTE can be found with respect to previous ADC(3) \cite{Mew14} and CASPT2 \cite{Gui17,Sot21} studies.

For 4-nitropyridine-$N$-oxide, which is often employed as a solvatochromic probe, \cite{Lag96} literature data are limited except for a dedicated theoretical investigation of the main band, \cite{Bud16b} and our recent contribution focusing 
on the CT EES. \cite{Loo21a}

Concerning phenolate, chosen for its low-lying Rydberg EES, no previous theoretical work specifically addressed the VTEs, while measurements are performed in solution which complicates direct comparisons due to both the significant stabilization of 
both the state and variations of EES ordering.

For pyranone (4$H$-pyran-4-one), accurate measurements and CCSD simulations of vibronic couplings are available for the lowest singlet and triplet EES,\cite{Ses19,Par23} but there is a lack of works dedicated to higher states. 
Our VTEs are approximately 0.5 eV larger than the measured 0-0 energies, which is acceptable for $\npi$ excitations involving a significant electronic reorganization. The singlet-triplet gap attains 0.18 eV with our TBE and 0.14 eV for the 
measured 0-0 energies.  \cite{Ses19,Par23} 

\begin{figure*}[htp]
  \includegraphics[width=0.8\linewidth]{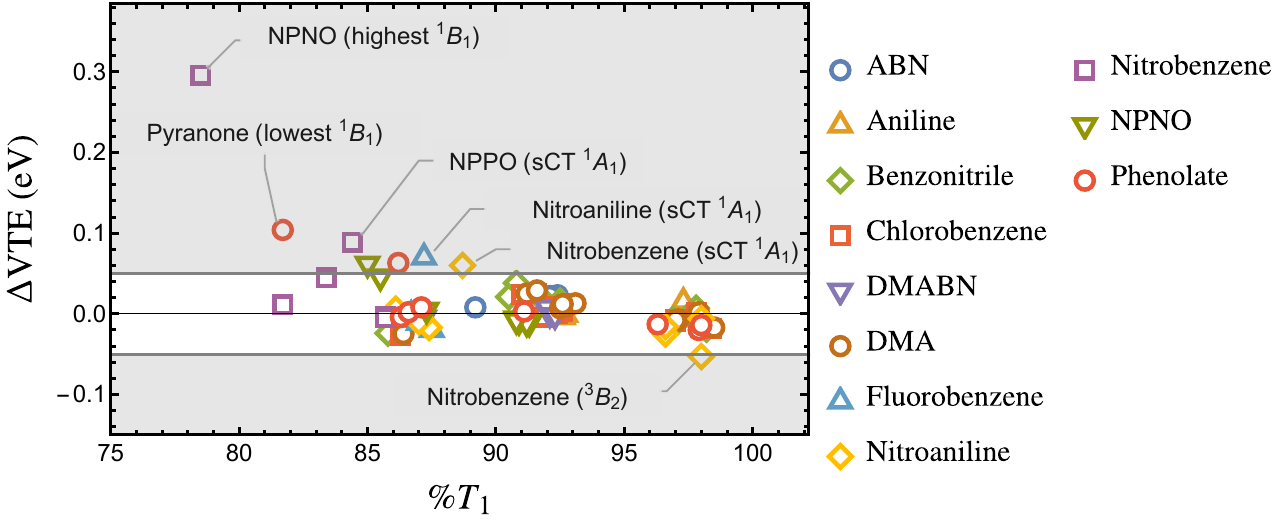}
  \caption{Difference between the CCSDT and CC3 VTEs, $\Delta$VTE (in eV), as a function of the percentage of single excitation involved in the transition ($\%T_1$) computed at the CC3/aug-cc-pVDZ level. The largest 
  available basis set is considered to determine the difference in VTEs. 
  The grey areas indicate errors larger than 0.05 eV. Selected outliers are specifically labeled.}
  \label{Fig-2}
\end{figure*}

Before going to the benchmark, let us discuss the quality of the TBE listed in Tables \ref{Table-1} and \ref{Table-2}.   It is important to note that, except for five EES (four in NPNO and one in pyranone), the single excitation character ($\%T_1$) 
for all considered transitions is at least  85\%\ for singlet and 95\%\ for triplet states. This result \emph{a posteriori} legitimates the use of CC methods to define the TBEs, as a recent investigation confirmed that large corrections 
from quadruples (beyond CCSDT) are expected only for EES with $\%T_1 < 85\%$. \cite{Loo22} Moreover, for transitions dominated by single excitations ($\%T_1 > 85\%$), CASPT2 and NEVPT2 tend to deliver significantly larger errors 
than CC3. \cite{Sar22} In Figure \ref{Fig-2}, the CCSDT correction, computed through Eqs.~\ref{eq-t-pop} or \ref{eq-t-dz}, is shown for the full set of data as a function of $\%T_1$. While the magnitude of this correction may not be a foolproof 
indicator of the significance of corrections introduced by quadruples, \cite{Loo22} it is evident that the CC3 and CCSDT VTEs are essentially equivalent for the vast majority of EES, providing further confirmation of the 
reliability of our results.

However, Figure \ref{Fig-2} reveals a few outliers that deserve specific discussions. Notably, for the three transitions exhibiting a substantial CT character in the nitro-bearing derivatives, CC3 tends to underestimate the VTEs, 
a phenomenon previously reported. \cite{Koz20,Loo21a} This is  why for all sCT EES listed in Tables \ref{Table-1} and \ref{Table-2}, we opted for CCSDT-3 rather than CC3 in Eqs.~\ref{eq-t-pop} and \ref{eq-t-dz}, as explained 
in Section \ref{sec:Comp}. The $^1B_1$ transition in pyranone, characterized by a $\%T_1$ value of 81.7\%, exhibits a notable CCSDT correction of $+0.104$ eV.  Fortunately, for this specific EES, CC4 calculations could be 
achieved (see below).  The most challenging EES, however, is the highest considered singlet state of NPNO, where a large difference of $+0.296$ eV is observed between the CC3 and CCSDT VTEs. This discrepancy is atypical for 
a state with a $\%T_1$ of 78.5\%. Indeed,  the well-known $^1A_g$ transition in butadiene, featuring a similar $\%T_1$ value (75.1\%), only exhibits a $-0.073$ eV difference between the CCSDT and CC3 results. \cite{Ver21} For 
NPPO, we noticed that the significant double excitation character found with CC3 decreases with CCSDT, possibly due to state mixing with another EES of the same symmetry, which explains the unexpected approximately 
$+0.3$ eV upshift. In any case, the TBE listed in Table \ref{Table-2} for this EES of NPPO should be considered as a very rough estimate. Lastly, an intriguing correction just above the 0.05 eV threshold ($-0.053$ eV) is surprisingly 
found for the lowest triplet EES of nitrobenzene, the rationale for which remains enigmatic.

For the 14 VTEs where CC4 corrections could be obtained, our confidence in the accuracy of the TBEs is logically higher. A noteworthy case is the previously mentioned $^1B_1$ EES of pyranone, characterized by a relatively low 
$\%T_1$ value of 81.7\%. For this state, the difference between CC4 and CCSDT VTEs, calculated with the 6-31+G(d) atomic basis set, remains modest at $-0.035$ eV. The CC4 transition energy is bracketed by its CC3 and CCSDT counterparts
but is logically closer to the latter. A similar, yet somewhat unexpected, CC4 correction is observed for the $^1A_1$ EES of the same molecule ($-0.037$ eV), displaying a $\%T_1$ value of 86.6\%. In all other cases, the improvements brought 
by CC4 are essentially insignificant, strongly hinting at the quality of the CCSDT values.

\subsection{Benchmark}

\subsubsection{Absolute accuracy}

With these TBEs at our disposal, it is natural to benchmark ``lower-order'' wave function methods, which could potentially be applied to larger or less symmetric compounds. In the vast majority of cases, the identification of each 
EES is straightforward thanks to the symmetry, oscillator strength, and underlying composition of molecular orbital pairs.

\begin{figure*}[htp]
  \includegraphics[scale=.9,viewport=2.5cm 14.5cm 18.5cm 27.5cm,clip]{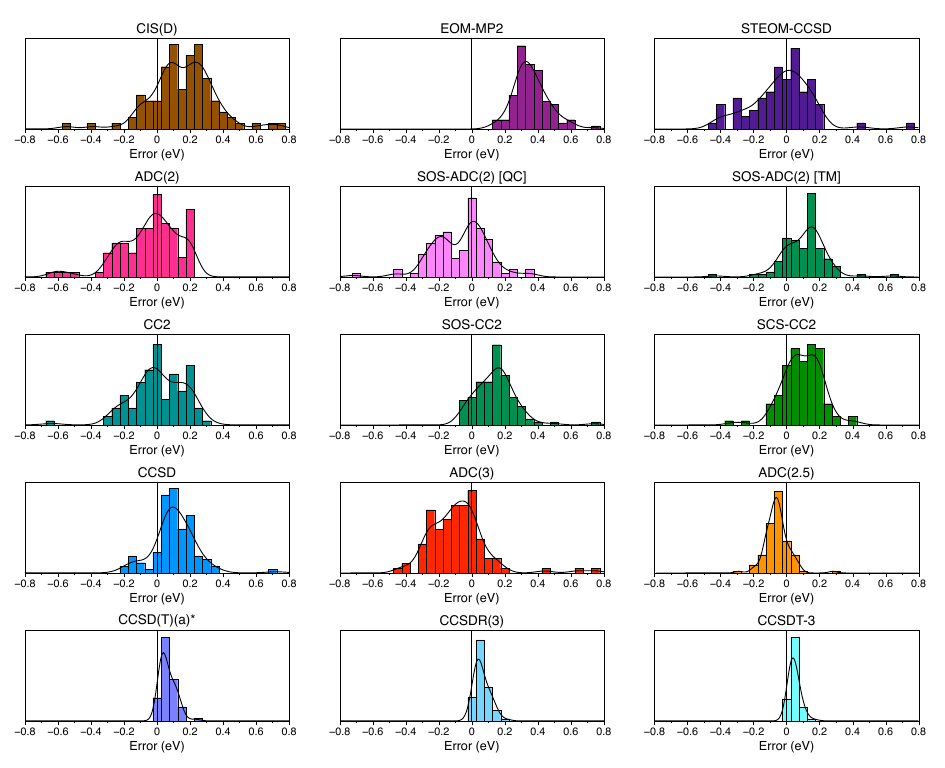}
  \caption{Distribution of the error (in eV) in VTEs (with respect to the TBEs) for various computational methods determined on the full set of molecules and states.}
  \label{Fig-3}
\end{figure*}

Table \ref{Table-3} presents the mean signed error (MSE), mean absolute error (MAE), standard deviation of the errors (SDE), root-mean-square error (RMSE), as well as the extremal positive and negative deviations 
[Max(+) and Max($-$)] for 15 popular and easily accessible wave function methods. The corresponding histograms of the error distribution are depicted in Figure \ref{Fig-3}. The analysis encompasses the full set of TBE listed in Tables \ref{Table-1} 
and \ref{Table-2}, excluding the problematic $B_1$ EES of NPPO for which a reliable estimate cannot be provided (as discussed earlier).

\begin{table}[htp]
\caption{Statistical data obtained for the full set of EES. All values are given in eV.}
\label{Table-3}
\begin{ruledtabular}
\begin{tabular}{lcccccc}
				&MSE	&MAE	&SDE	&RMSE	&Max(+)	&Max(-) \\
\hline
CIS(D)			&0.13	&0.21	&0.25	&0.28	&0.74	&-1.20 \\
EOM-MP2			&0.36	&0.36	&0.10	&0.37	&0.74	&0.13 \\
STEOM-CCSD		&-0.03	&0.14	&0.16	&0.19	&0.74	&-0.44 \\
ADC(2)			&-0.06	&0.15	&0.21	&0.22	&0.22	&-0.93 \\
SOS-ADC(2) [QC]	&-0.07	&0.13	&0.17	&0.18	&0.36	&-0.72 \\
SOS-ADC(2) [TM]	&0.11	&0.13	&0.13	&0.17	&0.64	&-0.43 \\
CC2				&-0.01	&0.14	&0.19	&0.19	&0.31	&-0.88 \\
SOS-CC2			&0.14	&0.15	&0.12	&0.19	&0.74	&-0.07 \\
SCS-CC2			&0.09	&0.12	&0.12	&0.15	&0.41	&-0.33 \\
CCSD			&0.10	&0.13	&0.13	&0.16	&0.72	&-0.20 \\
ADC(3)			&-0.06	&0.17	&0.24	&0.25	&0.97	&-0.44 \\
ADC(2.5)			&-0.06	&0.08	&0.07	&0.09	&0.28	&-0.29 \\
CCSD(T)(a)$^\star$	&0.06	&0.06	&0.05	&0.08	&0.23	&0.01 \\
CCSDR(3)		&0.06	&0.06 	&0.05	&0.07	&0.21	&-0.01 \\
CCSDT-3			&0.05	&0.05 	&0.03	&0.06	&0.17	&0.01 \\
\end{tabular}
\end{ruledtabular}
\end{table}

As anticipated from previous studies, \cite{Goe10a,Jac15b,Loo18a,Loo20a,Ver21,Loo21} CIS(D) yields poor results with relatively large MAE and SDE.  For the current homologous series, EOM-MP2 VTEs are 
excessively large, exhibiting a MAE of 0.36 eV (compared to 0.22 eV in the original QUEST database \cite{Ver21} and even smaller values for very compact compounds \cite{Kan17}). Despite these large errors, the EOM-MP2
discrepancies are systematic, resulting in a remarkably low SDE of 0.10 eV. This noteworthy outcome distinguishes EOM-MP2 with the smallest SDE among the tested methods presenting a favorable $\mathcal{O}(N^5)$ 
scaling with system size.

Consistently with previous findings, \cite{Hat05c,Win13,Har14,Jac15b,Kan17,Loo18a,Loo20a,Ver21,Loo21} ADC(2) and CC2 exhibit similar behavior, displaying error distributions centered around zero. Their 
MAE of approximately 0.15 eV and SDE of roughly 0.20 eV are characteristic of these two levels of theory. STEOM-CCSD, compared to CC2, showcases slightly improved accuracy, again agreeing with previous reports. 
\cite{Loo18a,Dut18,Loo20a,Loo20d,Loo21} We underline that, in STEOM-CCSD, some of the most challenging EES could not be fully converged (see Section \ref{sec:Comp}) and are, therefore, not considered in this 
assessment. This aspect might partially account for the observed outcome.

While the spin-scaling parameters of Q-Chem \cite{Kra13} were found more effective than their Turbomole counterparts \cite{Hel08} for ADC(2) in the QUEST database, \cite{Ver21} the conclusion is not as straightforward for the present set. Here, 
the former (latter) tends to underestimate (overestimate) the reference values.  The spin-scaled variants of CC2, SOS-CC2 and SCS-CC2, deliver nearly identical MAEs as the original CC2, but notably smaller SDEs. 
This latter observation aligns with previous findings, \cite{Goe10a,Jac15b,Taj20a,Loo20d,Ver21,Loo21} although unexpectedly, the MAE of CC2 and its two spin-scaled variants are essentially equivalent. Among the $\mathcal{O}(N^5)$ 
models reported in Table \ref{Table-3}, SCS-CC2 emerges as a promising compromise for the substituted phenyl compounds considered in this study.

The trends observed with SCS-CC2 are consistent with those of CCSD, showing a smaller SDE than both CC2 and ADC(2). 
However, this improvement comes at the cost of a notable blueshift of all transition energies. 
The overestimation trend associated with CCSD has been previously documented, \cite{Sch08,Car10,Wat13,Kan14,Kan17,Dut18,Ver21,Loo21} with the magnitude of overestimation appearing to depend on both the size of the system and its chemical nature.

ADC(3) does not yield a significant improvement over ADC(2), while ADC(2.5), the average between the two VTEs, proves to be the computationally cheapest scheme providing both MAE and SDE below the 0.10 eV threshold. 
This represents a noteworthy achievement for a $\mathcal{O}(N^6)$ composite approach proposed quite recently. \cite{Loo20b,Bau22}

The last three rows of Table \ref{Table-3} indicate that even more accurate values can be obtained with CC methods incorporating perturbative or iterative contributions from the triple excitations. As observed previously for bicyclic derivatives, \cite{Loo21} 
CCSD(T)(a)$^\star$ and CCSDR(3) exhibit extremely similar levels of accuracy, while the improvement brought by the iterative triples in CCSDT-3 is moderate for the present set of EES.

In Table \ref{Table-4}, we present MAEs determined for several EES subsets. It should be noted that we did not analyze CT excitations separately, given their limited presence in the set, comprising  7 instances only. 
Additionally, the $\npi$ statistics are derived from a relatively small set of 16 transitions across 4 compounds, including 3 with nitro substituents. 

\begin{table}[htp]
\caption{MAEs (in eV) associated with various subsets of EES.}
\label{Table-4}
\begin{ruledtabular}
\begin{tabular}{lcccccc}
				&Singlet	&Triplet	&Val. 	&Ryd.	&$\npi$	&$\ppi$\\
\hline
CIS(D)			&0.19	&0.26	&0.24	&0.14	&0.14	&0.27\\
EOM-MP2			&0.39	&0.30	&0.37	&0.33	&0.45	&0.35\\
STEOM-CCSD		&0.13	&0.16	&0.15	&0.13	&0.15	&0.15\\
ADC(2)			&0.15	&0.15	&0.15	&0.17	&0.27	&0.12\\
SOS-ADC(2) [QC]	&0.16	&0.08	&0.15	&0.09	&0.23	&0.13\\
SOS-ADC(2) [TM]	&0.13	&0.14	&0.12	&0.15	&0.11	&0.13\\
CC2				&0.13	&0.15	&0.11	&0.19	&0.09	&0.12\\
SOS-CC2			&0.13	&0.20	&0.17	&0.10	&0.30	&0.14\\
SCS-CC2			&0.09	&0.17	&0.14	&0.07	&0.19	&0.13\\
CCSD			&0.16	&0.09	&0.15	&0.10	&0.20	&0.14\\
ADC(3)			&0.14	&0.22	&0.18	&0.15	&0.19	&0.17\\
ADC(2.5)			&0.08	&0.08	&0.08	&0.07	&0.09	&0.07\\
CCSD(T)(a)$^\star$	&0.06	&		&0.08	&0.04	&0.14	&0.07\\
CCSDR(3)		&0.06	&		&0.07	&0.05	&0.13	&0.06\\
CCSDT-3			&0.05	&		&0.06	&0.04	&0.08	&0.05\\
\end{tabular}
\end{ruledtabular}
\end{table}

Table \ref{Table-4} highlights that two approaches, namely CCSD and SOS-ADC(2) [QC], exhibit significantly better performance for triplet than for singlet EES.  The opposite is found for ADC(3),  SOS-CC2,  SCS-CC2, and CIS(D).
This observation, except for CCSD, deviates from the results reported in QUEST, \cite{Ver21} implying that the relative performance for the two types of states may be contingent on the specific test molecules chosen, rather than 
being solely determined by the method under evaluation.

For most methods considered in the present analysis, the VTEs are slightly more accurate for Rydberg than for valence EES. The quality of Rydberg transitions appears to be more dependent on the basis set size than on the treatment of the electronic 
correlation, whereas the opposite holds true for valence transitions. A notable deviation from this trend is observed in the case of CC2, which is significantly poorer for Rydberg than valence EES, aligning with the conclusions drawn 
by Kannar, Tajti, and Szalay. \cite{Kan17}

In the case of $\npi$ transitions, errors are generally larger than those listed in QUEST, \cite{Ver21} indicating that these transitions pose a particular challenge for most models, likely due to the reasons outlined above, i.e., the
specific nature of the nitro compounds.

Notably, among all tested methods, only ADC(2.5) and CCSDT-3 deliver MAE smaller than 0.10 eV for all considered subsets. This emphasizes the notable accuracy of these two methods across a diverse range of EES.

\subsubsection{Chemical trends}

As discussed in Section \ref{sec:Intro}, the emphasis in EES calculations for practical applications often revolves around obtaining accurate chemical trends rather than absolute values. This is particularly pertinent when an experimental 
reference is available for certain compounds within a series. The selection of data of interest is contingent upon the specific application. For example, optimizing a dyeing application might require a focus on the most intense absorption, 
while many fluorophores might only necessitate consideration of the lowest EES. 

In this study, we have examined three specific parameters for which trends have been investigated: i) the VTEs to the first $^1B_2$ EES, which corresponds to the lowest $\ppi$ transition for most compounds 
in the series; ii) the position of the lowest $^3A_1$ EES, which often represents the lowest $\ppi$ triplet EES; and iii) the singlet-triplet gap (STG) considering the two lowest excited states ($S_1$-$T_1$ gap) for 
each method, regardless of their chemical nature and symmetry. In each case, we computed auxochromic shifts relative to the corresponding EES in benzene, the reference non-substituted compound. Our TBEs for the 
relevant singlet and triplet EES in benzene are 5.045 eV and 4.159 eV, respectively. \cite{Loo22}

\begin{figure*}[htp]
  \includegraphics[scale=.95,viewport=2.5cm 17.0cm 18.5cm 27.5cm,clip]{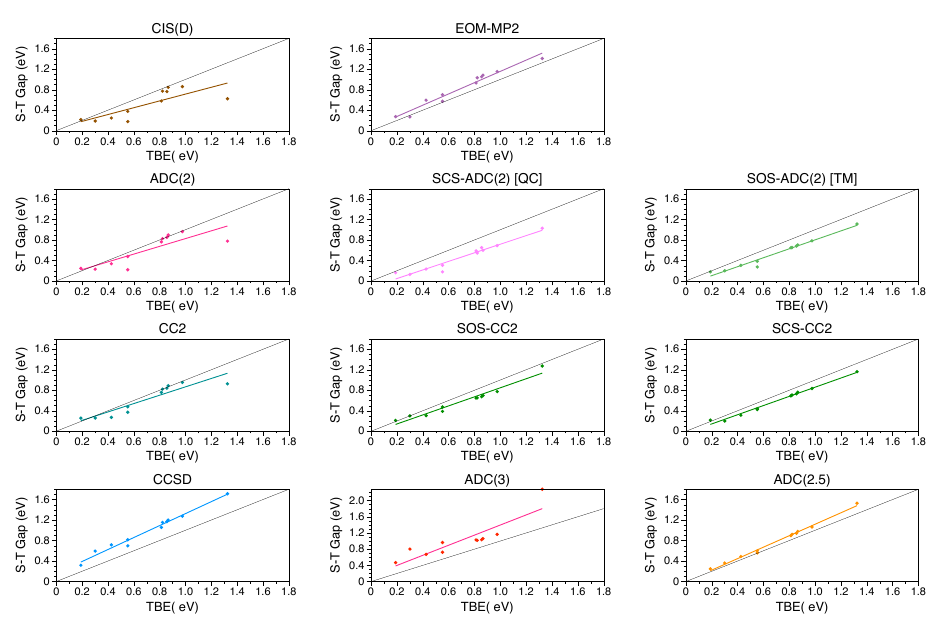}
  \caption{Correlation graphs between the STGs and the TBEs (both in eV) for various computational methods determined on the full set of molecules.}
  \label{Fig-4}
\end{figure*}

Our findings are summarized in Table \ref{Table-5}, where we present the MSE and MAE determined for the auxochromic effects. Additionally, we include the linear determination coefficient ($R^2$) obtained by comparing the 
shifts obtained with a given method to their TBE counterparts. As an illustrative example, correlation graphs for the STG can be found in Figure \ref{Fig-4}. It is important to note that due to missing data, we have not included 
STEOM-CCSD in this comparison, ensuring that our statistics are consistently based on the same number of transitions.  We also recall that, for the CC models including triples -- namely, CCSD(T)(a)$^\star$, CCSDR(3), and 
CCSDT-3 -- no implementation of triplet EES is available.

\begin{table*}[htp]
\caption{MSEs and MAEs (in eV) as well as linear determination coefficients ($R^2$) obtained when comparing the shifts with respect to benzene for the three properties discussed in the main text.}
\label{Table-5}
\begin{ruledtabular}
\begin{tabular}{lccccccccc}
				&\multicolumn{3}{c}{$^1B_2$ VTE}		&\multicolumn{3}{c}{$^1A_1$ VTE}		&\multicolumn{3}{c}{Singlet-triplet gap}			\\
										\cline{2-4}						\cline{5-7}						\cline{8-10}
				&MSE	&MAE	&$R^2$			&MSE	&MAE	&$R^2$			&MSE	&MAE	&$R^2$			\\
\hline
CIS(D)			&-0.05	&0.07	&0.995			&0.05	&0.06	&0.989			&-0.18	&0.19	&0.635\\
EOM-MP2			&0.06	&0.07	&0.987			&0.10	&0.10	&0.986			&-0.10	&0.10	&0.962\\
ADC(2)			&-0.18	&0.17	&0.953			&-0.08	&0.08	&0.994			&-0.14	&0.15	&0.707\\
SOS-ADC(2) [QC]	&0.03	&0.06	&0.984			&-0.03	&0.05	&0.991			&0.04	&0.07	&0.938\\
SOS-ADC(2) [TM]	&0.04	&0.05	&0.986			&0.01	&0.05	&0.988			&0.00	&0.05	&0.967\\
CC2				&-0.13	&0.12	&0.977			&-0.03	&0.04	&0.997			&-0.11	&0.12	&0.859\\
SOS-CC2			&0.07	&0.06	&0.984			&0.04	&0.04	&0.990			&0.06	&0.07	&0.950\\
SCS-CC2			&0.01	&0.02	&0.996			&0.02	&0.03	&0.995			&-0.01	&0.03	&0.987\\
CCSD			&0.03	&0.03	&0.992			&0.06	&0.06	&0.986			&-0.06	&0.08	&0.979\\
ADC(3)			&0.04	&0.05	&0.977			&-0.03	&0.04	&0.997			&0.14	&0.15	&0.791\\
ADC(2.5)			&-0.07	&0.07	&0.997			&-0.05	&0.05	&0.999			&-0.03	&0.05	&0.993\\
CCSD(T)(a)$^\star$	&0.00	&0.01	&0.999			&\\
CCSDR(3)		&-0.01	&0.02	&0.998			&\\	
CCSDT-3			&0.00	&0.01	&0.999			&\\
\end{tabular}
\end{ruledtabular}
\end{table*}

A notable positive observation from Table \ref{Table-5} is that most methods generally do provide accurate chemical trends, often with large to very large $R^2$ values, indicating a good correlation between calculated and reference shifts. 
However, a few exceptions can be identified.

For the first property, the change in the VTE of the lowest $^1B_2$ EES, CIS(D) stands out with a remarkably small MAE of 0.07 eV and an $R^2$ close to unity, despite its large SDE in Table \ref{Table-3}. This illustrates 
that a method may provide reasonable chemical trends for a specific case even if it exhibits rather poor performance for absolute energies. In contrast, both ADC(2) and CC2 show large deviations and relatively weak correlations 
for this auxochromic effect.  Opting for any of the spin-scaled approaches significantly improves their performance. The MSE and MAE of ADC(2.5) are similar to those in Table \ref{Table-3} for absolute energies, yet the correlation 
with the TBEs remains excellent.  The three CC schemes including triplets performance show very good performance.

For the second property, the transition energy to the lowest $^3A_1$ EES,  almost any of the evaluated wave function approaches (except perhaps EOM-MP2) produces highly satisfactory results. It appears that the evolution of the relative energy 
of this EES when adding chemical substituents to benzene, is easily captured by all wave function methods.

The most interesting case is probably the STG. We underline that, for some compounds, methods may disagree on the nature of the lowest singlet or triplet states, which introduces an additional aspect in the comparison, since in contrast to the
two former properties, the actual ordering of the EES provided by each method matters. As evidenced by the data gathered in Table \ref{Table-5} and depicted in Figure \ref{Fig-4}, CIS(D), ADC(2), and ADC(3) exhibit relatively poor correlations 
and, therefore, cannot be recommended for the STG of phenyl derivatives. Once again, the spin-scaling technique significantly improves the trends, although the absolute values of the STGs remain too small, especially with the two 
SOS-ADC(2) models. Interestingly for heptazine derivatives, which present inverted STG, the opposite was found, i.e., a superior performance of ADC(2) compared to its spin-scaled counterparts. \cite{Loo23a}

The STG values delivered by CCSD tend to be too large, which is a logical consequence of the larger (positive) errors observed for the singlet than for the triplet EES. Despite the substandard performance of ADC(2) and ADC(3), ADC(2.5) 
 delivers excellent estimates of the STG.

It is essential to acknowledge that the three simple tests presented above are not representative of all EES chemistry, and the results of Table \ref{Table-5} should not be directly extrapolated to other chromophoric units. Nevertheless, 
this comparison clearly illustrates that, except for the computationally expensive CC models that consistently perform well, lower-order methods can excel in predicting important chemical trends despite an overall modest performance 
(as seen in the case of CIS(D) for the $^1B_2$ and $^3A_1$ relative energies, despite a large overall MAE and SDE).

\section{Conclusions}

In summary, we have extended the QUEST database of accurate vertical excitation energies by adding 108 new VTEs obtained on 12 substituted six-membered rings, considering both donor and acceptor groups, as well as several push-pull configurations.
Except for one case where a notable difference between CC3 and CCSDT estimates was identified, we maintain confidence in the accuracy of the obtained TBEs. This confidence arises from the incorporation of CC4 corrections, the application 
of CCSDT-3 for CT states, or the manifestation of minor CC3-CCSDT discrepancies alongside prominent single-excitation characters.

Utilizing these TBEs, we conducted a thorough assessment of several second-order (ADC(2), CC2, CCSD, CIS(D), EOM-MP2, and STEOM-CCSD) and third-order (ADC(3), CCSDR(3), CCSD(T)(a)$^\star$, and CCSDT-3) wave function models.  
Crucially, our findings highlight the nuanced nature of method performance, revealing that some lower-order methods exhibit reasonable trends for some key EES in the present homologous series, despite large average errors and standard 
deviations when considering all EES. This underlines the necessity to taylor method selection based on the specific targeted application, as benchmarking solely on absolute transition energies may not offer a comprehensive understanding for 
applications involving similar compounds.

For the current set of compounds, taking into account all evaluated criteria, we suggest the adoption of ADC(2.5) and SCS-CC2 as accurate and reliable $\mathcal{O}(N^6)$ and $\mathcal{O}(N^5)$ approaches, respectively. Surpassing 
the performance of these models necessitates the use of CC approaches that incorporate triple excitations, albeit at the expense of a considerable increase in computational resources.

We are currently exploring additional chromophoric units to augment and diversify the content of the QUEST database.

\section*{Acknowledgements}
PFL thanks the European Research Council (ERC) under the European Union's Horizon 2020 research and innovation programme (grant agreement no.~863481) for financial support. DJ is indebted to the CCIPL/Glicid
computational center installed in Nantes for a generous allocation of computational time. 

\section*{Data availability statement}
The data that supports the findings of this study are available within the article and its supporting information.

\section*{Supporting Information}
Additional supporting information can be found online in the Supporting Information section at the end of this article.

\bibliography{biblio-new}

\end{document}